\documentstyle[psfig]{mn}
\def\ale{\mathrel{\hbox{\rlap{\hbox{\lower4pt\hbox{$\sim$}}}\hbox{$<$}}}}
\def\age{\mathrel{\hbox{\rlap{\hbox{\lower4pt\hbox{$\sim$}}}\hbox{$>$}}}}
 
\begin{document}

\title[The Redshift of Gamma-Ray Bursts] {Using STIS to find Gamma-Ray
Burst Redshifts}

\author[Bloom et al.]{Joshua S.~Bloom, Steinn Sigurdsson, Ralph
A.~M.~J.~Wijers, Omar Almaini, \cr Nial R.~Tanvir, and Rachel A.~Johnson 
\\ Institute of Astronomy, Madingley
Road, Cambridge, CB3 0HA, England \\
email: {\tt jsbloom@ast.cam.ac.uk}}
 
\date{}
 
\maketitle
 
\begin{abstract}

A recent spectrum of the optical afterglow of GRB 970508 suggests that
gamma-ray bursts (GRBs) are cosmological in origin and it is of
crucial importance to derive an accurate distance to each burst.  If
GRBs occur near their host galaxies ($<< 40$ kpc) then Lyman limit
absorption [$N({\rm HI}) \ge 1.6 \times 10^{17}$ cm$^{-2}$] should be
observable in roughly half the GRB afterglow spectra.  Here we outline
the methodology to obtain a redshift from the GRB afterglow spectrum
using the recently installed Space Telescope Imaging Spectrograph
(STIS) instrument onboard the {\it Hubble Space Telescope}.  A
low--resolution spectrum with the Multi-Anode Microchannel Array
(MAMA) detector gives complete spectral coverage over the wavelength
range 1570--3180 \AA (Near UV; NUV) and 1150--1740 \AA (Far UV; FUV).
Assuming a Target of Opportunity observation is conducted soon ($\ale
3$ weeks) after a bright burst, a relatively small integration time
($\sim 3$ orbits) would be sufficient to detect the Lyman limit over a
wide redshift range ($0.3 \ale z \ale 2.2$).  Detection (or
non--detection) of the Lyman limit, in concert with ground-based
observations of nearby galaxies and Mg II and C IV absorption lines,
should provide meaningful constraints on the relationship of GRBs to
galaxies.
\end{abstract}


\section{Introduction}

An optical source associated with gamma-ray burst (GRB) 970508 has
recently been detected (Bond 1997).  Absorption features seen in a
spectrum from the Keck 10m telescope indicate that it is at or beyond
$z=0.835$ and the lack of a prominent Lyman-$\alpha$ absorption
suggests $z \ale 2.1$ (Metzger et al.~1997). Presuming that the source
is indeed the optical afterglow of the burst, then GRBs have only just
been confirmed to be cosmological. As no emission lines have been
detected from the transient and it is unclear whether the burst is at
the redshift of the absorption system at $z=0.835$ (see \S 2), an
alternative way to get a limit on the redshift is by looking for a
Lyman-$\alpha$ forest or Lyman limit absorption.  Lyman limit
absorption arises from neutral hydrogen (HI) which is optically thick
to Lyman-continuum radiation for wavelengths $\lambda \le 912$ \AA\ in
the rest frame of the absorbing system.  Since both redshifted
Lyman-$\alpha$ absorption (1216 \AA) and the Lyman limit ($\lambda \le
912$ \AA) remain in the UV passband for $z \ale 2.2$, currently only
the Space Telescope Imaging Spectrograph (STIS) onboard the {\it
Hubble Space Telescope} (HST) can detect a Lyman limit. As expected
from theoretical models, both optical transients detected thus far
have started fading soon after discovery, and thus a STIS spectrum
must be obtained as soon as possible after the burst so as to maximise
signal--to--noise (S/N).

The distance inferred from a Lyman limit in the continuum would
provide knowledge of the luminosities and energies involved in the
explosion---two vital parameters for constraining GRB models.  In
section 2 we discuss the possibilities of the GRB afterglow undergoing
Lyman limit absorption. Then, in section 3 we discuss the
instrumentation and calculate the integration time required to infer
the redshift of a GRB.

\section{Lyman Limit Absorption Expectations}

Lyman limit absorption systems, which are generally HI clouds opaque
($\tau \age 1$) to the Lyman--continuum, are believed to concentrate
in the disk and halo of most galaxies. Steidel (1993) has found that
the density of Lyman limit systems is high enough out to impact
parameter of $\sim 40$ kpc in the galactic disk that continuum
radiation passing through the plane of the disk will always be
subjected to Lyman limit absorption.  Thus, if GRBs occur at offsets
$<< 40$ kpc from the centres of their host galaxy, it is expected that
roughly 50 percent (half in front of the disk, half behind the disk)
of GRB afterglow will have a Lyman limit break in the spectrum; this
Lyman limit will correspond to the precise redshift of the burst since
the limit system will be local to the GRB.

Intervening galaxies, not associated with the GRB but in the
line--of--sight of the afterglow, may also absorb the continuum; thus
a redshift inferred from a Lyman limit will not necessarily be the
redshift of the GRB. What effect will this have on the determination
of GRB redshifts?  Storrie-Lombardi et al.~(1994) survey QSO
absorption spectra and find that for any random line--of--sight, the
density of intervening Lyman limit systems is $N(z) \simeq 0.38 (1 +
z)^{1.04}$ for redshifts $z < 3.0$.  Thus at redshifts $z \age 1.5$ it
is expected that most GRB afterglows will be subjected to at least one
Lyman limit in their continuum that does not necessarily correspond to
the intrinsic redshift of the GRB.

\begin{figure}
\centerline{\psfig{file=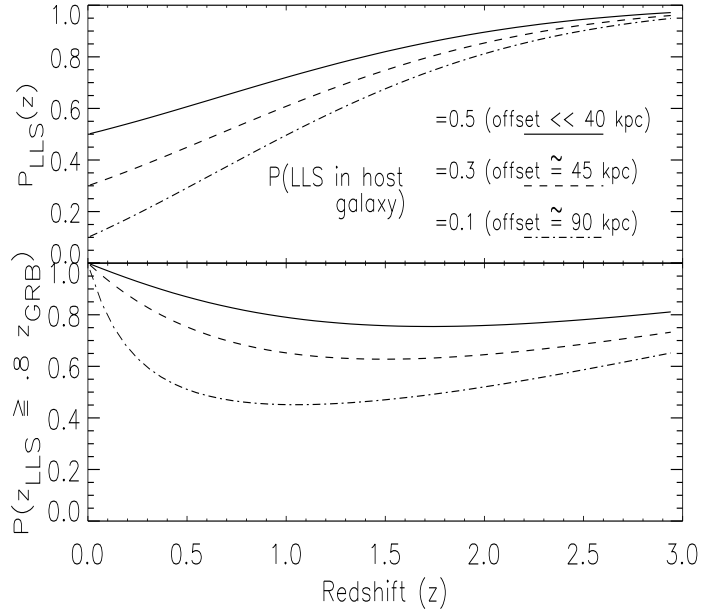,height=3.6in,width=3.6in}}
\caption{Top plot: The expected probability of the existence a Lyman
limit in the spectrum of a GRB afterglow as a function of redshift and
the offset scale length of GRBs from their host galaxy.  The
contribution to the total probability of a Lyman limit system (LLS)
from the host galaxy (0.5 solid; 0.3 dashed; 0.1 dot dashed) is
computed by assuming the GRBs occur randomly around the galaxy and the
impact parameter of the LLSs in the disk is $\sim 40$ kpc (Steidel 1993). The
contribution from intervening LLS (not associated with the GRB itself)
is calculated assuming the number density of LLS systems evolve as
$N(z) = 0.38(1+z)^{1.04}$ (Storrie-Lombardi et al.~1994). Bottom plot:
The probability, if a Lyman limit is present in the spectrum of the
afterglow, that the inferred redshift is accurate to within 20 percent
of the true redshift of the GRB afterglow itself. Not surprisingly at
low redshift, a distance inferred from the Lyman limit is probably the
true distance to the GRB since the expected number of LLSs from intervening
galaxies is small. See Appendix A for details.}
\label{fig:lls}
\end{figure}

If GRBs are ejected to distances comparable to the scale length of the
Lyman limit absorption systems in the disk, then the probability that
the host galaxy will absorb the spectrum shortward of the Lyman limit
is reduced.  The top portion of figure (\ref{fig:lls}) shows the
expected probability of the existence of a Lyman limit ($\tau \ge 1$)
in the spectrum of a GRB afterglow as a function of redshift and the
offset scale length of GRBs from their host galaxy.  The relationship
between the frequency of absorption from the host galaxy and the
offset scale is computed by assuming that absorption only occurs if
the GRB is seen through the 40 kpc absorbing disk and that the disk
has random viewing inclination.  As seen, the frequency of Lyman limit
absorption in the spectrum of GRBs at low redshifts ($z \ale 1$) may
be used to determine the intrinsic offset of GRBs from their host
galaxies since most Lyman limit absorption at low redshifts is
expected to come the host galaxy.
The bottom half of figure (\ref{fig:lls}) shows, as a function of
offset scale and redshift, the probability that the inferred redshift
is within 20 percent of the redshift of the GRB afterglow.  If GRBs
occur with about 60 kpc of their host galaxy (solid line), then more
than 60 percent of the redshifts inferred from the Lyman limit in the
spectrum will be a moderately accurate ($< 20$ percent) measure of the
redshift of the GRB.

Absorption from Mg II, with a galactic impact parameter of $\sim 50$
kpc (Bergeron et al.~1994), and C IV is expected, but not required, to
accompany a Lyman limit system (Storrie-Lombardi et al.~1994). Thus,
figure (\ref{fig:lls}) could also be seen as a prediction of the
frequency of absorption lines in GRB afterglows.  Indeed for GRB
970508, both Mg II and Fe II absorption was detected using Keck
(Metzger et al.~1997); Arav \& Hogg (1997) have found that the absence
of detectable C IV in the spectrum of the afterglow limits the
redshift of the GRB to $z \le 1.8$ (95 percent confidence).

\section{Observing Details}

The STIS CCD instrument onboard HST with a G230LB low--resolution
grating gives complete spectral coverage over the wavelength range
1685--3065 \AA. This would be sufficient to detect the Lyman limit
over the redshift range ($0.85 \ale z \ale 2.4$).  The preferred
instrument for obtaining redshifts from UV spectra, however, are the
Near UV (NUV) and Far UV (FUV) Multi-Anode Microchannel Array (MAMA)
detectors which is more sensitive and has much wider spectral range
(FUV: 1150--1736 \AA; NUV: 1570--3180 \AA) corresponding to a broad
redshift range ($0.26 \ale z \ale .9$ and $0.72 \ale z \ale 2.5$,
respectively).

\subsection{Signal--to--Noise (S/N) estimation}

Temporal fits to the broadband spectra of the optical transients GRB
970228 (Van Paradijs et al.~1997) and GRB 970508 (Bond 1997) indicate
that the GRB afterglow may follow a power-law both in flux and
spectral shape in optical wavelengths.  It should be noted that the
later transient had an observed rising phase of optical emission
(e.g.~Galama et al.~1997).
Although individual cases may vary, in calculating the expected signal
as a function of wavelength and time, we adopt the functional form of
the flux as:
\begin{equation}
F_\nu(t) = F_{t_0} \nu^{-\delta} (t - t_{0})^{\beta} 
\end{equation}
where $F_{t_0}$ properly normalises the spectrum at $t_0$, the time at
which the decay begins.  Wijers, Rees, \& M\'esz\'aros (1997) find
that the afterglow adequately fits the early light curve with $\delta = 0.8$ and
$\beta = -1.2$ for GRB 970228.  Our preliminary fits to the data from
GRB 970508 indicate that $\delta \simeq 0.8$ and $\beta \simeq -1.0$.
If the afterglow is observed as a Target of Opportunity $t_{\rm obs}
\ale 3$ weeks after the burst (e.g.~$U(t_{\rm obs}) \simeq 22.7$ for
GRB 970508), a redshift could be obtained in less than 10 HST orbits
(see fig.~[2]) using the either FUV or NUV MAMA detectors.
\begin{figure}
\centerline{\psfig{file=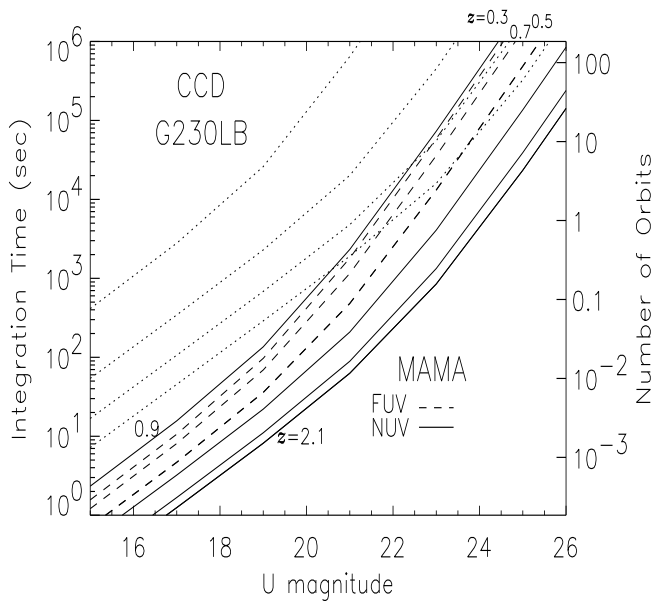,height=3.5in,width=3.5in}}
\caption{The expected integrating time required to achieve a S/N=3 in
a 100 \AA\ bin as a function of $U$ magnitude of the afterglow and
redshift of the Lyman limit source.  The MAMA detectors on STIS (NUV,
solid lines; FUV dashed) achieves a significantly better
signal--to--noise than the CCD (dotted lines).  For both MAMA NUV and
the CCD, the redshifts correspond to (left to right): $z=0.9$,
$z=1.2$, $z=1.5$, and $z=2.1$. The FUV curves correspond to (left to
right): $z=0.3$, $z=0.7$, and $z=0.5$. Also shown on the right
vertical axis is the number of HST orbits required assuming the object
is observable for 90 minutes of each orbit.  The signal--to--noise is
calculated for the indicated grating with a $52 \times$ 0.2 arcsec
slit and assuming average Zodiacal light, bright earth level.}
\label{fig:int}
\end{figure} 

Unless {\it HST} is able to observe the afterglow of a GRB while it is
still bright ($U \ale 20$), detection of damped Lyman $\alpha$
absorption at low redshift ($z \ale 1.5$) will require a very large
integration time on STIS.  However, a significant detection of a Lyman
limit requires far less S/N per unit wavelength and thus improves the
chance of determining a redshift of GRBs from fewer orbits.  Although
the Lyman break occurs at shorter wavelengths ($\lambda \le 912 (1 +
z)$ \AA) than the Lyman $\alpha$ forest ($\lambda \simeq 1216
(1+z)$\AA) where the detectors are less sensitive, the distinct
cut-off of this spectral feature is unambiguous and does not require
good spectral resolution, making this the most effective and clearcut
way to limit the redshift of faint sources.

Figure (\ref{fig:int}) shows the expected integration time required to
achieve a S/N=3 in a 100 \AA\ bin as a function of $U$ magnitude of
the afterglow and redshift of the Lyman limit source for both the CCD
and MAMA detector on STIS\footnotemark\footnotetext{Calculated using
the STIS Spectroscopic Exposure Time Calculator located at {\tt
http://www.stsci.edu/}.}.  As seen in the figure, if the afterglow is
observed while it is still reasonably bright ($U \ale 21$), detection
of a redshift $z \age 0.3$ would require less than 1 HST orbit.
\begin{figure}
\centerline{\psfig{file=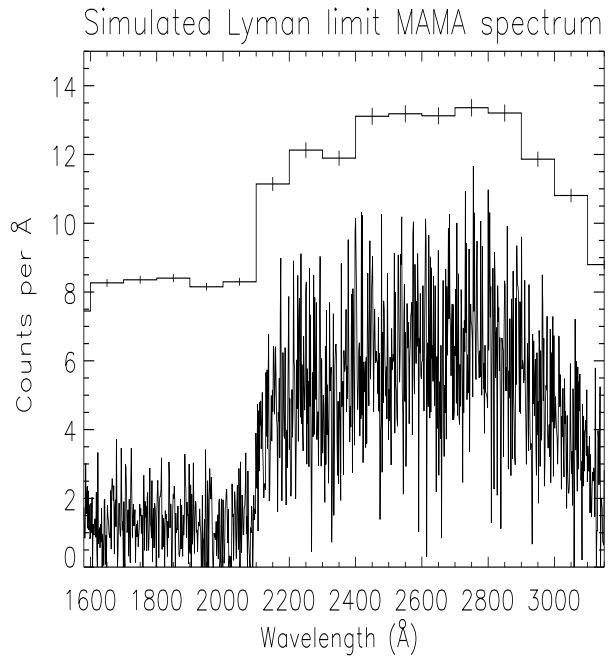,height=3.6in,width=3.6in}}
\caption{Simulated afterglow spectrum with $\sim 1.5$ HST orbits (5400
sec). The spectrum is computed for the MAMA $52 \times$ 0.2 arcsec G230L
slit on STIS; the source is taken to be a magnitude $U = 21.0$ and the
spectral shape is $\delta=1$ (eq.~1).  We have removed from the
spectrum the narrow OII line (2470 \AA) expected from geocoronal
emission. The Lyman limit is clearly detected at $\lambda \simeq 2100$
\AA\ corresponding to a redshift $z \simeq 1.3$.  The histogram
overlay, with estimated error bars, is the renormalised counts per 100
\AA\ with arbitrary offset.}
\label{fig:mama}
\end{figure}
Figure (\ref{fig:mama}) shows a simulated spectrum of an afterglow
obtained with MAMA with $\sim 1.5$ HST orbits (5400 sec) where the
source is a magnitude $U = 21.0$ and the spectral shape is $\delta=1$
(eq.~1).  We have removed the contribution of the source to the
observed spectrum for wavelengths $\lambda \le 2098$ \AA\
corresponding to a Lyman limit at a maximum redshift of $z=1.3$.  The
clear break in the spectrum implies that a fairly accurate redshift is
obtainable.

\section{Discussion}

 In general, given the efficiency of STIS and the spectral shape of
the afterglow, the S/N increases with higher redshift (see
fig.~[\ref{fig:int}]) for the NUV MAMA detector.  If the burst occurs
at a redshift $z \age 1.5$ then, with only a few orbits, one could
determine the position of the Lyman break well enough to determine its
redshift to an accuracy of better than 5 percent; it may even be
possible to detect a Lyman-$\alpha$ forest at high $z$ with only a few
orbits.  As it is easier to detect a redshift for bursts that
originate from a higher $z$ with the NUV MAMA detector, the absence of
a Lyman limit in the spectrum would, in general, always provide an
upper limit to the redshift.  From figure (\ref{fig:int}), it is clear
that use of the MAMA detectors onboard the HST, given their very low
internal countrates, are preferable to the CCD.

The disadvantage of using HST to infer a redshift is the difficulty of
altering its scheduled observations even on the timescale of
weeks. Certainly the advantage is that there are very few available
ground-based telescopes that can resolve Mg II and C IV absorption
lines from faint objects; even then, such telescopes may not have the
proper viewing conditions to the source and detected lines may place
different limits in the redshift.  In addition, there may be important
emission (e.g.~lines) in the UV spectrum of GRBs that would not be
observable from the ground.

As the ensemble of optical counterparts begins to grow in size, it
will be possible to infer both their redshift distribution and the
distribution of GRBs with respect to observable galaxies by noting the
frequency of absorption lines and detected Lyman limit (see
fig.~\ref{fig:lls}).  If no Lyman limit is detected in the spectra of
the optical transients (especially at low-redshifts), the conclusion
would be that either GRBs do not originate near galaxies ($\ale 300$)
kpc or they are Galactic in origin.

\section{Conclusions}

A precise  redshift of a gamma-ray burst would greatly further the
field by providing an accurate understanding of the energies involved
in the burst.  Assuming the extrapolation of the spectral index
($\delta$) from the optical to UV passband is correct, we find that a
MAMA detector observation of the bright afterglow of a GRB by STIS
over $\sim 1$ orbit could reveal a Lyman limit in the spectrum and
hence, if the GRB occurs behind a region of neutral hydrogen of its
host galaxy, provide a direct redshift to a gamma-ray burst.  Given
that Lyman limit absorption may come from galaxies along the
line--of--sight, we predict the expected frequency of Lyman limits in
the spectrum as a function of GRB redshift and the distance from
their host galaxy. 

\section*{Acknowledgements}

The authors thank Richard McMahon, Max Pettini and Phillip Outram for
insightful comments.  This paper has also benefitted from
conversations at the workshop, ``Recent Developments Towards
Understanding Gamma-Ray Bursts'' in Elba, Italy.

\section*{Appendix A: LLS Probability}

A Lyman limit system (LLS) is defined as the region of space for which
the optical depth ($\tau$) to the Lyman-continuum is greater than one.
Most Lyman limit systems are believed to be clouds of HI with a column
density $N({\rm HI}) \ge 1.6 \times 10^{17}$ cm$^{2}$.  As depicted in
the top half of figure 1, the probability as a function of redshift
that a spectrum of a GRB will contain at least one such LLS is
\begin{eqnarray*}
{\rm P}( \ge 1 {\rm ~LLS}) &=& 1 - {\rm Poisson}\left[0~{\rm
	LLS}~|~m(z_{\rm min},z) \right] \\
	&& \times~{\rm P}({\rm No~LLS~from~host~Galaxy})
\end{eqnarray*}
where ${\rm Poisson}\left[0~{\rm LLS} | m(z_{\rm min},z) \right]$ is
the Poisson probability of no intervening LLS between the GRB source
(at redshift $z$) and the observer given the expected number of LLSs:
\begin{equation}
m(z_{\rm min},z) = \int_{z_{\rm min}}^{z} N(z') dz'. \nonumber
\end{equation}
The minimum redshift in which a LLS could be detected is ${z_{\rm
min}}$ and $N(z)$ is the number density of LLSs per unit redshift.

The probability that the redshift inferred from an observed Lyman limit in a
GRB afterglow spectrum is at least $x$ times the redshift of the GRB
itself is given as
\begin{equation}
{\rm P}(z_{\rm LLS} \ge x \times z | \ge 1~{\rm LLS}) = 
\frac{{\rm P(A~}|{\rm ~B})~{\rm P(B)}}{\rm P(A)}~~{\rm where,}
\end{equation}
\begin{equation}
{\rm P(A~}|{\rm ~B}) = {\rm P}( \ge 1~{\rm LLS} |~{\rm
LLS~redshift} \ge x \times z) = 1,
\end{equation}
\begin{equation}
{\rm P(A)} = {\rm P}( \ge 1 {\rm ~LLS}), 
\end{equation}
\begin{eqnarray}
{\rm P(B)} &=& {\rm P}( z_{\rm LLS} \ge x \times z) \\
           &=&  1 - {\rm Poission}\left[0~{\rm LLS}~|~m(x \times z,z)
           \right] \nonumber
	 \\
	&& \times~{\rm P}({\rm No~LLS~from~host~Galaxy}). \nonumber
\end{eqnarray}
This probability is depicted in fig.~1 for $x=0.8$.


\begin{thebibliography}{99}

\bibitem[]{}Arav, N.~and Hogg, D.~W.~1997, ApJ, submitted.

\bibitem[]{}Bergeron, J.~et al.~1994, ApJ, 436, 33.

\bibitem[]{}Bond, H.~E.~1997, IAU Circular 6654.

\bibitem[]{}Galama, T.~J., et al.~1997, IAU Circular, 6655.

\bibitem[]{}Metzger, M.~R., Djorgovski, S.~G., Steidel, C.~C.,
Kulkarni, S.~R., Adelberger, K.~L., and Frail, D.~A.~1997, IAU
Circular 6655.

\bibitem[]{}Steidel, C.~C., and Sargent, W.~L.~1992, ApJS, 80,1.

\bibitem[]{}Steidel, C.~{\it The Environment and Evolution of
Galaxies, eds. J. M. Schull \& H. A. Thornson, Dodrecht: Kluwer} {\bf
263}.

\bibitem[]{}Storrie-Lombardi, L.~J., McMahon, R.~G., Irwin, M.~J.,
Hazard, C.~1994, ApJ, 427, L13.

\bibitem[]{}Van Paradijs, J.~et al.~1997, Nature, 386, 686. 


\end{thebibliography}
\end{document}